\documentclass[12pt]{article}
 \usepackage{amsmath,amssymb}

 \newcommand{\be}{\begin{equation}}
 \newcommand{\ee}{\end{equation}}
 \newcommand{\bl}{\begin{equation}\begin{array}{ll}}
 \newcommand{\el}{\end{array}\end{equation}}
 \newcommand{\bll}{\begin{equation}\begin{array}{lll}}
 \newcommand{\bdm}{\begin{displaymath}}
 \newcommand{\edm}{\end{displaymath}}
 \def\bea{\begin{eqnarray}}
 \def\eea{\end{eqnarray}}
 \def\barr{\begin{array}}
 \def\earr{\end{array}}
 \newcommand{\bean}{\begin{eqnarray}}
 \newcommand{\eean}{\end{eqnarray}}
\def\p{\partial}

\def\f{\varphi}
\def\ve{\varepsilon}
\def\ep{\epsilon}
 
 \def\de{\delta}
 \def\la{\lambda}
 \def\La{\Lambda}
 \def\al{\alpha}
 \def\ga{\gamma}

\def\half{\frac{1}{2}}

\def\2third{\frac{2}{3}}
\def\4third{\frac{4}{3}}
\def\3quart{\frac{3}{4}}

\def\pr{\prime}

\def\bg{\bar{g}}

\def\bl{\bar{l}}

\def\bl{\bar{l}}

\def\cA{{\cal A}}

\def\cF{{\cal F}}

\def\da{\dot{a}}

\def\doq{\dot{q}}

\def\dvphi{\dot{\f}}

\def\ddpsi{\ddot{\psi}}
\def\dchi{\dot{\chi}}
\def\dxi{\dot{\xi}}
\def\ddxi{\ddot{\xi}}

\def\dpsi{\dot{\psi}}
\def\dal{\dot{\alpha}}

\def\deta{\dot{\eta}}

\def\dF{\dot{F}}
\def\ddF{\ddot{F}}

\def\hve{\hat{\varepsilon}}

\def\ha{\hat{a}}

\def\hh{\hat{h}}

\def\hA{\hat{A}}

\def\hU{\hat{U}}

\def\hrho{\hat{\rho}}

\def\Lag{{\cal L}}

\def\prpsi{\psi^{\prime}}
\def\pprpsi{\psi^{\prime\prime}}
\def\pprxi{\xi^{\prime\prime}}

\def\prPhi{\Phi^{\prime}}
\def\prchi{\chi^{\prime}}
\def\prho{\rho^{\prime}}
\def\preta{\eta^{\prime}}
\def\prh{h^{\prime}}
\def\prH{H^{\prime}}
\def\prG{G^{\prime}}
\def\prF{F^{\prime}}

\def\ppz{z^{\prime\prime}}
\def\pz{z^{\prime}}

\usepackage{amsmath}

\DeclareMathOperator{\arccosh}{arcCosh}

\begin{document}
\raggedbottom

  \title{{\bf Integrals of equations for cosmological and static reductions in generalized theories of gravity
 }}

\author{A.T.~Filippov \thanks{Alexandre.Filippov@jinr.ru} \\
{\small \it {$^+$ Joint Institute for Nuclear Research, Dubna, Moscow
 Region RU-141980} }}

\maketitle

\begin{abstract}
  We consider the dilaton  gravity models derived by reductions of generalized theories of gravity and study one-dimensional dynamical systems simultaneously describing cosmological and static states in any gauge. Our approach is fully applicable to studying static and cosmological solutions in multidimensional theories and also in general one-dimensional dilaton - scalaron gravity models. We here focus on general and global properties of the models, on seeking integrals, and on analyzing the structure of the solution space. We propose some new ideas in this direction and derive new classes of integrals and new integrable models.

\end{abstract}

\section{Introduction}
 Current observational data strongly suggest that Einstein's gravity must be modified. The combination of data on dark energy and the growing evidence for inflation have generated a wide spectrum of such modifications. Superstring and supergravity ideas suggested natural modifications, but in view of the serious mathematical problems of the current string theory, strict unambiguous predictions about concrete modifications of gravity are not yet available. Moreover, the phenomenon of dark energy was not predicted by string theory, and its origin in the stringy framework proved rather difficult to uncover and understand. The problem of dark energy in string theory seems very deep and is related to many other complex issues of quantum cosmology, but it also led to some beautiful and exciting  speculations, like eternal inflation and the multiverse. On the other hand, if we first try to find a natural place for dark energy in classical cosmological models, which are almost inevitably essentially nonlinear and non-integrable, then we had best return to recalling the origin of general relativity and seek some options abandoned or not found by its creators.

   Therefore, simpler modifications of gravity that affect only the gravitational sector are also popular. In essence, these modifications reduce to the standard Einstein gravity supplemented by some number of scalar bosons (the first example of such a modification was the old Jordan –- Brans –- Dicke
 theory). The main problem with this approach is that the origin of these scalar bosons is
                    not clear,
 and there is no theoretical principle governing their coupling to gravity. Of course, there exist some
 phenomenological and theoretical constraints, but the spectrum of these models is generally too wide.\footnote{Restricting consideration to homogenous cosmologies, one can find that in dimensionally reduced supergravity theory there may emerge massless scalar bosons that couple to gravity only, see, e.g., \cite{Lidsey}.}
  The modification proposed and  studied in \cite{ATF} - \cite{ATF13}
    satisfies some principles of geometric nature based on Einstein's idea (1923)\footnote{References to Einstein's papers as well as to many papers of other authors, which are related to the subject of this report, can be found  in our publications \cite{ATF} - \cite{VDATF2}. } to formulate gravity theory
   in a non-Riemannian space with a symmetric connection determined by   a special variational principle involving a \emph{`geometric' Lagrangian}. This Lagrangian is  assumed to be a function of the generalized Ricci curvature tensor and of other fundamental tensors, and is varied in the connection coefficients. A new interpretation and  generalization of this approach was developed in \cite{ATFn} - \cite{ATFs} for arbitrary space-time dimension $D$.

  The connection coefficients define symmetric and antisymmetric parts ($s_{ij}$ and $a_{ij}$) of the Ricci tensor and a new vector $a_i$.
   Assuming that  there are no dimensional fundamental constants in the pure geometry (except the speed of light relating space to time) we choose geometric Lagrangians giving a dimensionless geometric action. The geometric variational principle puts further bounds on the geometry and, in particular, relates $a_i$ to $a_{ij}$.  To define a metric tensor we must introduce a dimensional constant. We then can find a \emph{physical Lagrangian} depending on this dimensional constant and on some dimensionless parameters. The theory thus obtained supplements standard general relativity with dark energy
   (the  cosmological term, in the limit  $a_i = a_{ij} =0$), a neutral massive (or tachyonic) vector field proportional to $a_i$ (a vecton), and after dimensional reductions to $D=4$,
   with $D-4$  massive (or tachyonic) scalar fields.

  The most natural density of this sort in any dimension is
  the square root of $\det(s_{ij} + \bl a_{ij})$, where $\bl$
 is a number.\footnote{Einstein used Eddington's scalar density $\sqrt{|\det r_{ij}|}\,$, where $r_{ij} \equiv s_{ij} + a_{ij}$,  as the Lagrangian.}
  The effective physical Lagrangian is the sum
 of the standard Einstein term, the vecton mass  term,
 and a term proportional to $\det(g_{ij} + l f_{ij})$
 to the power $\nu \equiv 1/(D-2)$, where  $g_{ij}$ and  $f_{ij}$
 are the metric and the vecton field tensors \emph{conjugate}
 to $s_{ij}$ and $a_{ij}\,$,\footnote{This unusual construction
 introduced by A.~Einstein is described and generalized in
 \cite{ATF} - \cite{ATF13}.}
 and $l$ is a parameter related to  $\bl$.
 The last term has a dimensional multiplier, which in the limit of a small field $f_{ij}$ produces the cosmological constant. For $D=4$, we therefore have the term first introduced by
 Einstein but now usually called the Born - Infeld or \emph{brane} Lagrangian. For $D=3$, we have the Einstein - Proca theory, which is very interesting for studies of nontrivial space topologies.

  Here, we consider the simplest geometric Lagrangian,
   \be
 \label{1.1}
 {\Lag}_{\rm geom} = \sqrt{-\det(s_{ij} + \bl a_{ij})}\,
 \equiv \, \sqrt{-\Delta_s}  \,,
  \ee
 where the minus sign is taken because $\det(s_{ij})<0$
 (due to the local Lorentz invariance) and we
 naturally assume that the same holds for
 $\det(s_{ij} + \bl a_{ij})$ (to reproduce Einstein's general relativity with the cosmological constant in the limit
 $\bl \rightarrow 0$). Following the steps in \cite{ATFn} or using results described in \cite{ATF13} we can derive the
 corresponding \emph{physical Lagrangian}
  \be
 \label{1.2}
 \Lag_{\rm phys} =  \sqrt{-g} \,
 \biggl[ -2 \Lambda \,[\det(\delta_i^j +
 l f_i^j)]^{\nu} +  R(g) -
 m^2 \, g^{ij} a_i a_j \biggr] \,, \qquad
   \nu \equiv 1/(D-2) \,,
 \ee
  which should be varied with respect to the metric and the
  vector field; $m^2$ is a parameter depending on the
  chosen model for affine geometry and on $D$
  (see \cite{ATF} - \cite{ATFs}). This parameter can be
  positive or negative and we often use notation $m^2 \equiv \mu$.
  When the vecton field vanishes,  we have the standard Einstein gravity  with the cosmological constant. For dimensional reductions from $D \geq 5$ to $D=4$, we can obtain the Lagrangian describing
 the vecton $a_i$, $f_{ij} \sim \p_i a_j - \p_j a_i$
 and $(D-4)$ scalar fields $a_k, \, k = 4,..,D$.
  We note that Lagrangian (\ref{1.2}) is bilinear in the vecton field, for $D=3$, and gives the three-dimensional gravity with the cosmological term in the approximation $a_i = 0$.

  To compactify this report we only give an overview of main points.
 First, consider a rather general Lagrangian in the
 $D$-dimensional \emph{spherically symmetric} case
 ($x^0 = t$, $x^1 = r$):
 \be
 ds_D^2 = ds_2^2 + ds_{D-2}^2 =
 g_{ij}\, dx^i\, dx^j \,+ \,
 {\f}^{2\nu} \, d\Omega_{D-2}^2 \, ,
  \label{2.1}
   \ee
  where $\nu \equiv (D-2)^{-1}$.
  The standard spherical reduction
  of (\ref{1.2}) gives the effective Lagrangian\footnote{
  We suppose that the fields $\f$, $a_i$ are dimensionless while
  $[t] = [r] = \textrm{L}$ and thus $[f_{ij}] = \textrm{L}^{-1}$,
  $[R] = [k_{\nu}] = [X] = [m^2] = \textrm{L}^{-2}$. For more details on our dimensions see \cite{ATF13}.}
  the first three terms of which describe the standard spherically
  reduced  Einstein gravity:
  \be
 \label{2.2}
 \Lag^{(2)}_D =  \sqrt{-g} \,  \biggl[\f R(g) +
 k_\nu \, \f^{1-2\nu} + W(\f)\, (\nabla \f)^2 +
  X(\f, \textbf{f}^{\,2}) - m^2 \f \, \textbf{a}^2
 \biggr] \,.
 \ee
 Here $a_i(t,r)$ has only two non-vanishing components
 $a_0, a_1$, $f_{ij}$ has just one independent component
 $f_{01} = a_{0,1} - a_{1,0}$; the other notations are:
 $\textbf{a}^2 \equiv a_i a^i\,\equiv g^{ij} a_i a_j\,$,
 $\textbf{f}^{\,2} \equiv f_{ij} f^{ij}\,$,
 $k_\nu \equiv k(D-2)(D-3)$, $W(\f) = (1-\nu) /\f$
 and, finally,
 \be
 \label{2.3}
 X(\f, \textbf{f}^{\,2}) \equiv
  -2 \Lambda \f \,\bigg[1 +
  \half \lambda^2 \textbf{f}^{\,2}\,\biggr]^{\,\nu} \,,
 \ee
    where, the parameter $\lambda$ is related to
    dimensionless parameter $l$ in (\ref{1.2}) but
    $[\la] = \textrm{L}$ .

 Sometimes, it is convenient to transform away the dilaton
 kinetic term by using the Weyl transformation, which in our
 case is the following ($w^{\pr}(\f) / w(\f) = W(\f)$):
 \be
 g_{ij} = \hat{g}_{ij} \, w^{-1}(\f) ,\,\,\,\,\,
 w(\f) = \f^{1-\nu}, \,\,\,\,\,\,
 \textbf{f}^{\,2} = w^2 \,\hat{\textbf{f}}^{\,2} \,,\,\,\,\,\,
 \textbf{a}^2 \,=\, w \, \hat{\textbf{a}\,}^2 \,.
 \label{2.4}
 \ee
 Applying this transformation to (\ref{2.2}) and omitting the
 hats we find in Weyl's frame that
  \be
 \label{2.5}
  \Lag^{(2)}_D \,\mapsto \, {\hat{\Lag}}^{(2)}_D  =  \sqrt{-g} \,
 \biggl[\f R(g) +  k_\nu \, \f^{-\nu} -
 2 \Lambda \f^{\nu} \,\bigg(1 +
  \half \lambda^2 \f^{2(1-\nu)}
  \textbf{f}^{\,2}\,\biggr)^{\,\nu}  -
  m^2 \f \, \textbf{a}^2 \biggr] \,.
 \ee
   When $D=3$ we have $\nu = 1$, $k_\nu =0$, Weyl's transformation
 is trivial and the Lagrangian is
  \be
 \Lag^{(2)}_{3}  =  \sqrt{-g} \, \f \biggl[ R(g) -
  2 \Lambda -  \lambda^2 \Lambda  \, \textbf{f}^{\,2} -
  m^2 \, \textbf{a}^2 \biggr] \,.
  \label{2.6}
 \ee

 These two-dimensional reductions are essentially simpler than
 their parent higher dimensional theories.
 In particular, we show that the massive vecton field theory
 can be transformed into a dilaton - scalaron gravity model
  (DSG) which is easier to analyze.
 Unfortunately, these DSG models and  their further
 reductions to dimension one (static and cosmological reductions)
 are also essentially non-integrable. It is well
 known that the massless case, being a pure dilaton gravity,
 is classically integrable. Having this in mind,  we will look for additional integrals of motion in similar reduced vecton theories which we transform into scalaron dilaton gravity models.

   The structure of the two-dimensional theory allows one
  to find some integrable classes using
  simplifying assumptions about their potentials.
  For some multi-exponential potentials and constant (`minimal')
  coupling of scalars, there exist integrable systems related
  to Liouville and Toda-Liouville ones (see \cite{ATF1} - \cite{VDATF2}). The pure Liouville case
   was completely solved.  For the Toda-Liouville, it is
   difficult to find exact analytical solutions of the
   two-dimensional constraints,
   even in the simplest $u_1 \oplus su_2$ case.
   This problem is easily solved in the one-dimensional
   (static or cosmological) reduction.
        Unfortunately, even the one-dimensional
   cosmological reduction of the pure scalaron theory
      is not integrable and thus we concentrate
   on searching for approximate potentials
    that allow us to find a wide enough class of analytic
   solutions to reconstruct exact ones by iterations.

 \section{Cosmological and static reductions}
 In this report, we consider only the reductions
 of the two-dimensional theory to static and cosmological equations ignoring one-dimensional waves studied in our previous work
 \cite{ATF2} - \cite{VDATF2}. The simplest reduced gauge fixed equations can be directly  derived by supposing that in the two-dimensional light-cone equations the fields $h, \f, q, \psi$ depend on one variable,
  which we denote $\tau \equiv (u+v)$. For the cosmological solutions
  this variable is identified with the time variable, $\tau = t$,
  while for the static states, including black holes, it is
  the space variable, $\tau = r$. The only difference between the
  cosmological and static solutions is in the sign of the metric, $h_{\rm c} > 0$,  $h_{\rm s} < 0$.

  In our study of black holes and cosmologies we use
  the more general diagonal  metric,
  \be
  ds_2^2 = e^{2\alpha (t, r)} dr^2  - e^{2\gamma (t, r)} dt^2 \,.
   \label{a.1}
 \ee
  Then, the static and cosmological reductions of our two-dimensional vecton theory (\ref{2.5}) supplemented with the standard scalar term $V(\f, \psi) + Z(\f) \nabla {\psi}^2$ can be presented by the Lagrangian
  (taking, for the moment, $Z(\f) = -\f$)
    \be
  \ep {\cal L}_{\rm v}^{(1)} = e^{\ep (\alpha - \gamma)} \f
  \biggl[\dpsi^2 - 2 \dal_{\ep} \, {\dvphi \over \f} -
  (1-\nu) \biggl({\dvphi \over \f}\biggr)^2 \biggr]
   - e^{\ep (\gamma-\alpha)} \mu \,\f \, a^2_{\ep} \,+\,
  \ep \, e^{\alpha + \gamma} \, \biggl[ V + X(\textbf{f}^{\,2})
  \biggr] \,.
  \label{a.2}
  \ee
   Here we omit a possible dependence of $V$ and $X$
   on $\f$ and $\psi$, denote $Z_a \equiv -\mu \f \equiv
   m^2$, and $\ep = \pm\,$.
   All the fields depend on $\tau = t$ ($\ep = +$)
   or on $\tau  = r$ ($\ep = -$). Finally,
   \bdm a_{+} = \,a_1\, (\tau)\,, \quad  a_{-} = \,a_0 \,(\tau) \,,
   \quad \alpha_{+} \equiv \alpha\,, \quad \alpha_{-} \equiv \gamma \,,
   \quad \dal_{\ep} = {d \over d\tau}\, \alpha_{\ep} \,, \quad
   \da = {d \over d\tau} \,a \,.
   \edm

   We see that the cosmological and static Lagrangians essentially
   coincide, the only difference being in the sign of the potential
   term and of the metric exponents as well.
   As the kinetic term depends on $\dal_{\ep}$,
   the multiplier of the kinetic term,  $l_{\ep} \equiv \exp(\alpha_{-\ep})$,
   is a Lagrange multiplier, varying of which produces the constraint equation that
   which is equivalent to vanishing of the Hamiltonian. In view of the  implicit
   dependence of  $\textbf{f}^{\,2}$ on $l_{\ep}\,$, it is much more convenient
   to first employ the canonical formulation and then identify the
   proper Lagrange multiplier. Omitting simple details given in \cite{ATF13}, we only give the final result.

   Introducing the Hamiltonian formulation for general theory (\ref{a.2})
   we can apply to it the elementary canonical transformation,
    \be
    p_a \Rightarrow -2q\,, \quad  a \Rightarrow p/2 \,, \qquad
    X_{\rm eff}(p_a) \Rightarrow X_{\rm eff} (-2q)\,,
    \label{5a}
    \ee
    and then derive the corresponding new Lagrangian
       \be
  \ep {\cal L}_q^{(1)} = l^{-1}_{\ep} \,
  \biggl[\f\, \dpsi^2 - 2 \dal_{\ep} \, \dvphi -
  (1-\nu) \, {\dvphi^2 \over \f} \,+\,
    {\doq^2  \over m^2 \f} \biggr]  +\,
   l_{\ep} \, \ep \,e^{2\alpha_{\ep}}\, U(\f, \psi, q) \,,
  \label{a.7}
  \ee
    where  $l_{\ep} \equiv\, \exp(\alpha_{-\ep} - \alpha_{+\ep})$ and
    $U = V(\f, \psi) + X_{\rm eff} (-2q)$. The detailed derivation of this \emph{transformation of the vecton into the scalaron}, explanation of notation, and discussions of the analytic approximate expressions of the potential for arbitrary $D$ can be found in \cite{ATF13}.

  This form is more natural than (\ref{a.2}). First, the dependence on the Lagrange multiplier  $l_{\ep}\,$ is simple and standard,
 the kinetic part is quadratic in generalized velocities and can be
 made diagonal by a redefinition of the Lagrange multiplier and
 velocities. In addition, we are free to make a convenient gauge choice and to choose the Weyl frame. For example, by making the shift
 $ \alpha_{\ep} \Rightarrow  \alpha_{\ep} -(1-\nu) \ln \sqrt{\f} $
  and redefining the potential by
 $U \Rightarrow \, \f^{\nu - 1} U$
 we remove the third term in (\ref{a.7}) and obtain the Lagrangian in
 Weyl's frame.
   Then we can redefine $l_{\ep} \f \equiv \bl_{\ep}$,
  introduce the new field $\xi \equiv \,\f^2$ and
  finally rewrite (\ref{a.7}) in a simpler form,
  \be
  \ep {\cal L}_q^{(1)} = \bl^{-1}_{\ep}\,
  [\,\xi \dpsi^2 +\, m^{-2} \doq^2   - \dxi \, \dal_{\ep} \,] \,+\,
   \bl_{\ep} \, \ep \,e^{2\alpha_{\ep}}\, \xi^{\nu/2 -1} \,
   U(\sqrt{\xi}\,, q, \psi) \,.
  \label{a.8}
  \ee
  Before applying it to studies of cosmologies
  and horizons in the scalaron theory we discuss the
  effective scalaron potential, corresponding to $X$-potential
   (\ref{2.3}) in more detail.

  Note that the scalaron kinetic term $\sim \doq^2 $ is independent of $D$ while the potential $U$ is simple only for $D=3$, $D=4$.
  In the Weyl frame (see (\ref{2.6}) and (\ref{2.5}), respectively),
   it is easy to derive the effective potentials
   $U / w(\f) \equiv \, \hU$:
    \be
     U(\f\,, q) =  \hU(\f,\, q) =
   -2\Lambda \f \,[\,1 + q^2 / 4\lambda^2 \Lambda^2 \,\f^2 \,],
   \qquad  D=3 \,,
      \label{3.14}
  \ee
  \be
   U(\f\,, q) =  \sqrt{\f} \, \hU(\f\,, q) = -2\Lambda \f \,
   [\, 1 + q^2 / \lambda^2 \Lambda^2 \f^2 \,]^{\half}
  \,+\, 2k\,, \qquad D=4 \,.
   \label{3.13}
  \ee
  The general effective potential in Lagrangian (\ref{a.8}) can be written as
  \be
  U_{e} (\f, x) \equiv \, \xi^{\nu/2 -1} \, U(\sqrt{\xi}\,, q, \psi)
  = \,\f^{\nu -2} U =\, - 2 \La \f^{\nu -1}\, v_{\nu}\, (x) \,+\,
  k_{\nu}\, \f^{-(1 + \nu)} \,,
  \label{a.15}
  \ee
  where $x \equiv q / (-2\nu \lambda \Lambda \,\f) \,,$ and
  $v_{\nu}(x)$ is monotonic concave function having simple
  expansions %%%for $x \ll 1$, $x \gg 1$:
  \be
   v_{\nu}(x) \,=\, 1 + \nu x^2 + O(x^4)\,; \qquad
   v_{\nu}(x) \,=\, 2\nu x \biggl[1 +
   {1-\nu \over 2\nu}  x^{-\sigma} + O(x^{-2\sigma}) \biggr] \,.
    \label{a.17}
  \ee

   With such a simple and regular potential $U_e\,$,
   one might expect that at least
  qualitative behavior of the solutions of the theory (\ref{a.8})
  could be analyzed for small and large values of $x$.
  This is true if the theory is integrable. But it is probably not
  integrable, even in the simplest $D=3$ case, when $v_1(x) = 1 + x^2$,
  $k_1 = 0$ and thus $U_e = -2\La (1+x^2)$ is linear in
  $q^2/\xi$. The form of the potential signals that there must exist at least one additional integral beyond the Hamiltonian constraint, and it was derived in \cite{ATF13}. The existence of a third integral, which should allow us to integrate the scalaron model, is doubtful. We still hope to find either a reasonable approximation for the potential or to treat the exact systems by use of approximate, asymptotic, and qualitative (topological) methods. In next Sections we attempt at presenting a draft panorama of old and new integrals in reasonably general dilaton gravity coupled to scalars, and briefly describe a few simple, intuitive approaches to search for new integrals.

   \section{Integrals and integrability in simple cases}
 Here we consider a general DGS with one scalar $\psi$
 that may be a standard  field or the scalaron. The general Weyl-frame Lagrangian can be written as
  \be
 {\cal L}_{\rm dgs}^{(2)} = \sqrt{-g}\, \biggl[ \f R
 + Z(\f)(\nabla \psi)^2
 + V(\f,\psi) \biggr] \,
  \label{a.24}
  \ee
  (note that, in Lagrangians, we usually write $+V$ instead of the standard $-V$). For the scalaron we have $\psi = q$, $Z = Z_q = -1 /(m^2 \f)$ and the potentials are given above. For the standard scalar $Z_{\psi} \sim -\f$, but the results presented below are applicable to more general $Z$-functions. In our notation,
 negative signs of $Z$ correspond to positive kinetic energies of
 the scalar fields but our classical consideration is fully applicable
 to both signs. The general model (\ref{a.24}) with a general
 potential $V$  is not integrable in any sense. One of the strongest  obstructions to integrability is the dependence of  $Z$ on $\f$, and the  usual simplifying assumption is that the $Z$-functions are independent of $\f$. With this restriction, there exists a special class of `multi - exponential' potentials, for which the DGS theories with   any number
  of scalar fields can be reduced to the Toda - Liouville
  systems  and exactly solved.\footnote{
  This class includes all previously considered integrable
  two-dimensional DGS, which are reviewed in \cite{ATF2}.  The first DGS of the Liouville type (`bi Liouville'), which generalizes
  the so-called Jackiw and CGHS models, was proposed and solved in paper \cite{ATF1}, the results of which were essentially generalized in \cite{ATFp} and \cite{ATF13}.}
  For their  static - cosmological reductions,
    analytic solutions were explicitly derived. Here we try to expand this class of the models.
  It is well known that for the constant field $\psi \equiv \psi_0$ the two-dimensional theory can be exactly solved with any potential $V(\f) \equiv V(\f, \psi_0)$. In fact it degenerates to a one-dimensional theory because the dilaton and the metric satisfy the D'Alembert equation and thus depend on one variable $\tau = a(u) + b(v)$. Thus the complete solution can be derived by solving elementary one-dimensional equations defined by Lagrangian (\ref{a.7}) with constant $\psi$ and $q$.

  A more general \emph{static-cosmological reduction} of the two-dimensional theory (\ref{a.24}) is:
  \be
  \ep {\cal L}^{(1)} = - l^{-1}_{\ep} \,
  \bigl[Z(\f)\, \dpsi^2 + 2 \dal_{\ep} \, \dvphi \, \bigr] +
     l_{\ep} \, \ep \,e^{2\alpha_{\ep}}\, V(\f, \psi) \,,
  \label{a.7a}
  \ee
  where $V$ is the Weyl-frame potential and $h \equiv \ep \,e^{2\alpha_{\ep}}$ can be identified with the Weyl-frame metric if we choose the gauge $l_{\ep} \,= 1$, which we call the LC gauge. In this gauge the equations of motion are most directly reduced to the parent  two-dimensional dilaton gravity in the
  light-cone ($u,v$) coordinates. Also useful is the Hamilton gauge, in which $l_{\ep} Z^{-1}(\f)\, \equiv \bl_{\ep}\, =1$; using in addition the new variable $\xi$ defined by $d\,\xi \equiv Z^{-1}(\f)\, d\f$ we have:
  \be
  \ep {\cal L}^{(1)} = - \bl^{-1}_{\ep} \,
  \bigl[\dpsi^2 + 2 \dal_{\ep} \, \dxi \,\bigr] +
     \bl_{\ep} \, \ep \,e^{2\alpha_{\ep}}\, U (\xi, \psi) \,,
  \label{a.7b}
  \ee
  where $U(\xi, \psi) \equiv Z(\f) V(\f, \psi)$. This gauge is especially convenient when there are many scalar fields with the same $Z(\f) = -\f$. If the potential is the sum of linear exponents of the scalar fields, there is a class of explicitly integrable models including Toda-Liouville dilaton gravity theories (see \cite{ATF1}-\cite{VDATF2}). Other gauge choices used in the theory of black holes and in cosmological models may also be exploited in our dynamical formulation, but they are usually less convenient in the context of our search for integrals of dynamical systems.

  Let us first write the dynamical equations in the \emph{Hamilton gauge}:
  \be
   \ddxi + h\,U = 0\,, \quad  2\,\ddpsi + h\,U_{\psi} = 0\,, \quad
   \ddF + h\,U_{\xi} = 0 \,;
   \qquad \dpsi^2 + \,\dF \dxi + h\,U = 0\,.
  \label{ad1}
  \ee
   Here the last equation is the Hamiltonian constraint,
   $F \equiv \ln |h| \equiv \, 2 \al_{\ep}\,$, and the lower indices $\f$,  $\psi$ denote the corresponding partial derivatives. Our first approach to integrability of this system was based on taking linear combinations of the equations,
   \be
  c_1\,\ddxi \,+\,c_2 \,\ddpsi \,+\, c_3 \,\ddF \,+\, \ep e^F \,
  [\,c_1 \,U \,+\, c_2 \,U_{\psi}/2 \,+\, c_3 \,U_{\xi}\,] = 0\,.
  \label{ad2}
  \ee
  If for a given potential $U$ the expression in brackets vanishes, we find an integral which is linear in momenta. We can also find a general solution of the partial differential equation for $U$ giving corresponding integrals. For multi-exponential potentials we can instead try to construct, with the aid of (\ref{ad2}), the Liouville or Toda equations choosing different $c_n$. This approach, first proposed in \cite{ATF1}, was applied to constructing an integrable ("bi-Liouville") two-dimensional DGS with $U_{\psi} = 0$.\footnote{
  In \cite{ATF1}, $Z$ was constant but here we can take arbitrary $Z(\f)$ as it is included in $U$.}
  Taking $c_3 = 1$, $c_2 = 0$, $c_1 = \pm \la_1$ we find two Liouville equations for
  $F_{\pm} \equiv (F \,\pm \,\la_1 \xi)$ if the potential $U$ satisfy two equations,
  $[U_{\xi} \,\pm\, \la_1 \,U]\, \exp(\mp \,\la_1 \xi) =\, g_{\pm}$. These conditions are satisfied by the simplest bi-exponential potential,
  \bdm
  2 \la_1 U \,=\, g_{+} \,\exp (\la_1 \xi) \,-\, g_{-}\,\exp (-\la_1 \xi)\,,
  \edm
   while $F_{\pm}$ satisfy the Liouville equations and the correspondent integrals
  \be
  \ddF_{\pm} \,+\, \ep \, g_{\pm}\,\exp{F_{\pm}} \,=\,0 \,; \quad
  \dF_{\pm}^2 \,+\, 2 \ep \,g_{\pm} \,\exp{F_{\pm}} \,=\,C_{\pm}\,.
  \label{ad3}
  \ee
 This gives the complete solution of the problem:
 $4 \la_1 \,\dpsi^2 = C$ may be taken as the third integral, and the constraint (\ref{ad1}) is satisfied if $C + C_{+} - C_{-} =0\,$.

    Note that if we tried a more general potential, with
   $-\la_1$ replaced by $\la_2$, we would immediately find that the necessary condition for the existence of two integrals is
    $\la_2 = -\la_1\,$. A similar solvable model is given by
   one-dimensional reduction (\ref{a.7b}) of theory (\ref{a.24}). Taking
   $V = g_1 \exp(2\la_1 \psi) \,+\, g_2 \exp(2\la_2 \psi)$,
   $Z = -\f \equiv -e^{-\xi}\,$ and $\psi_1 \equiv (F - \xi)/2$,
   $\psi_2 \equiv (F + \xi)/2$, $\psi_3 \equiv \psi$ we first rewrite the potential term in (\ref{a.7b}) as
   \bdm
   h U = \ep \,e^F Z\,V = - (g_1\, e^{q_1} + g_2 \,e^{q_2})\,, \qquad
   q_i \equiv 2 (\psi_i + \la_i \,\psi_3) \,,
   \edm
  then find that the kinetic term is diagonal in $\doq_i^2$ if
  $\la_1 \la_2 = 1$, and see that the Lagrangian is
    \bdm
  {\cal L}  = -\bl^{-1} [- \mu_1\,\doq_1^2 + \,\mu_2\,\doq_2^2 + \,\dpsi_3^2\,]
   - \bl \,(g_1\, e^{q_1} + g_2 \,e^{q_2})\,; \qquad
  4 \mu_1 = -(1 - \la_1^2)^{-1}\,, \quad \mu_2 = -\la_1^2 \,\mu_1 \,.
   \edm
  The solution is obtained as in the previous case but the formulation can obviously be generalized to any number of scalar fields. If there is $N-2$ scalar fields with the same coupling to the dilaton, $Z(\f)$, the \emph{multi-exponential} models are defined by the Lagrangian
  \be
  {\cal L} = \,-\bl^{-1} \,
  \sum_1^N \ve_n \dpsi^2_n \,-      \,
     \bl\, \,\sum_1^N g_n \,e^{q_n} \,,\qquad
     q_n \equiv \sum_1^N \psi_m a_{mn} \,,
  \label{ad4}
   \ee
   where $\psi_1, \psi_2$ were defined above, $\ve_1 = -1$,
   $\ve_n = 1, \,n>1$. The properties of this multi-exponential theory depend on the symmetric matrix $\hA \equiv \ha^T \hve \,\ha$, where $\hve_{mn} \equiv \ve_m \de_{mn}$. If $\hA$ is diagonal, we have  $N$-Liouville model, which was directly solved also in the two-dimensional case. Such models were met in simple compactifications of supergravity theories \cite{Lidsey}.

     More complex compactifications may lead to integrable models that are related to Toda systems.
   In fact, any matrix $A_{mn}$, which is a direct sum of
  a diagonal ($L\times L$) matrix $\gamma_n^{-1} \delta_{mn}$
  and of a symmetric matrix $\bar{A}_{mn}$,
  can be represented in the form $\hA \equiv \ha^T \hve \,\ha$
   if the sum of $\gamma_n^{-1}$ is a certain function of the matrix elements $\bar{A}_{mn}\,$, see \cite{VDATF2}.
  Then, if $\bar{A}_{mn}$ is a direct sum of Cartan matrices, the differential equations for the $\psi$-functions can be reduced to
  $L$ independent Liouville (Toda $\cA_1$) equations and the
  higher-rank Toda system.  We call it the \emph{Toda-Liouville} system,  applications of which to the dilaton gravity was discussed in \cite{VDATF2}. The class of multi-exponential models is the main source of completely integrable dilaton gravity theories with many degrees of freedom. Most of them can be analytically solved in one-dimensional case while the two-dimensional $N$-Liouville and the simplest Toda-Liouville theories were classically solved in two dimensions.

  The one-dimensional exact solutions can easily be quantized
  (at least, formally) as their Hamiltonians split into the sum of the Liouville Hamiltonians $p^2 + 2g e^q$, where we suppress the indices of $p_n, q_n$ and factors $\ve_n$ in (\ref{ad4}). Introducing the new canonical variables
  \bdm
  P \equiv \sqrt{p^2 + 2ge^q}\,, \qquad
  Q \equiv -\arccosh (1 + p^2 g^{-1} e^{-q}) \,,
  \edm
  we find the complete solution of the $N$-Liouville theory:
  \be
  H =  \sum \ve_n P_n^2 \,,  \quad
  Q_n = P_n (\tau - \tau_n) \,;  \qquad
  \exp{(-q_n)} = g_n P_n^{-2} \cosh^2\, [P_n (\tau - \tau_n )/2\,] \,,
   \label{ad5}
 \ee
 where we `recovered' sign factors enabling to satisfy the constraint
 $H = 0$. In fact, we were a bit sloppy in writing these simple formulae,  and one must be more accurate, especially when quantizing this apparently trivial theory. Anyway, this is a constrained (gauge) theory and should be quantized as such. In more complex, especially, in non-integrable gauge models discussed below quantizing is more tricky and will not be discussed here.

    With small number of scalar fields, there may exist other integrals,
  not related to the Toda-Liouville class. The simplest integrals emerge when $U(\xi, \f)$ is independent either of $\xi$ or on $\psi$.
  Remember that $U \equiv Z(\f) V(\xi,  \psi) /w(\f)$, where $w$ is the Weyl transition function defined above and thus $V$ can be nontrivial even when $U_{\xi} = U_{\psi} = 0$. If the potential is a constant the system is obviously explicitly integrable. The integrals can exist also when the potential satisfies some weaker restriction; we first discus the potential $U = g\,\xi +\, v(\psi)$. In the simplest case, when $g=0$, the theory can be explicitly solved if, in addition, $v(\psi)$ is a simple function.  The main simplification is that $F = a \tau$; assuming that $v = \bg \,\psi^p$, we find the following rather tractable equations to be discussed in a moment:
    \be
  \pprpsi (F) = -\,\gamma p \,e^F \psi^{\,p-1} \,, \quad
  \pprxi (F) = -\,\gamma \,e^F \psi^p \,;
   \qquad \gamma \equiv \ep \bg\,a^{-2} \,.
  \label{ad6}
  \ee

  The simplest model is $U = g$. Then, with $N-2$ scalar fields $\psi_n$  ($3 \leq  n  \leq N$),  we have $N$ integrals. Supposing that all the scalars have the same  $|Z(\f)|$ we find
      \bdm
  \dF = C_1 \equiv a\,, \quad \dxi + (g/a)\, h  = C_2\,,
  \quad \dpsi_n =C_n\,, \quad  C_1 C_2 \,=\, \sum \ep_n \, C_n^2 \,,
  \edm
   where $\ep_n = +1$ for the normal field and $\ep_n = -1$ for the phantom (ghost) field $\psi_n\,$, and thus
     \be
  \xi - \xi_0 = -\hh + \ln |\hh|^{\,\delta}\,; \qquad \hh \equiv h/h_0 \,,
  \quad h_0 \equiv C_1^{\,2}/g \,, \quad
  \delta \equiv C_2/C_1 \,=\, \sum \ep_n \, (C_n / C_1)^2  \,.
  \label{ad7}
  \ee
  The first equation is what we call the  \emph{portrait} of the integrable DGS system.
  We consider only the cosmological part of it which looks like the portrait of a more realistic cosmological system. It is instructive to draw it in the $(\f, \hh)$-plane supposing that $\xi- \xi_0 = \ln \f$.
  One then can see that the \emph{separatrix} $\delta = 0$ describes the solution with one \emph{horizon} while the two other correspond to $|\delta| = 1$.
  In addition, the portrait has singularities. To get the physics portrait in the $(\f, \hh)$-plane one should first identify all possible singularities in this plane: the saddle point of the horizon, $(1,0)$, the nodes $(0,0), (0,1),(\infty,0), (\infty, 1)$ and the most interesting node $(1/e, 1)$ -- the cross-point of all cosmological solutions. This portrait topologically resembles the part of one derived in \cite{ATF1}. However, even these parts are globally inequivalent because the structure of their characteristic nodes is  not equivalent under differentiable topological mappings.

  For the linear and quadratic potentials $v(\psi)$ we can derive  explicit analytic solutions. If $p = 1$, we find a very simple solution of equations (\ref{ad6}) and of the constraint:
  \bdm
  \psi = -\gamma \,(e^F + C_1 F + C_2)\,, \quad
  \xi - \xi_0 \,=\, \gamma^2 \,[\,e^{2F}/\,4 \,+ (C_2 - 2C_1)\,e^F +  C_1 Fe^F - C_1^2 \,F/\,2\,]\,.
  \edm
   These expressions are similar to the simplest solutions discussed above and we do not discuss them in detail. One can see that the horizon  emerges when $C_1 = 0$ and
  $F \rightarrow -\infty \,$: then $h \equiv \ep e^F \rightarrow 0$ while $\xi$ is finite, $\xi \rightarrow \xi_0\,$. The topological portrait, $\xi(h)$, of this system is somewhat more complex than the previous one and we cannot discuss it here.

  In the quadratic case, $p = 2$, the $\psi$-equation is linear and can be solved in terms of the Bessel functions,
  $\psi = Z_0(2\sqrt{2 \gamma e^F})$. The $\xi(F)$ for the horizon separatrix,
  \bdm
  \xi - \xi_0 \,=\, \,\psi_0^2 \,\sum_{n=0}^{\infty}
  d_n \,(n+1)^{-2} \,(-2 \gamma \,e^F )^{\,n+1} \,, \qquad
  d_n \equiv \sum_{m=0}^n \,[\,m! \,(n-m)!\,] \,,
  \edm
  generalizes the expressions above while the complete portrait contains similar entire function multiplied by powers of $F$ and depending of an additional parameter $\delta$.
  It follows that the qualitative properties of this model do not radically differ from the previous cases.

   When $v = \bg^p$ and $p \neq 0,1,2$, we may expect much richer picture. If $p$ is real, $p \neq 0,1,2$, Eq.(\ref{ad6}) is called the \emph{Emden - Fowler equation}. It was studied in great detail, especially, for integer and rational values of $p$, see e.g. \cite{Bel}.  As is well known,  it has a very important but simple \emph{enveloping} solution,
  \be
   \psi_p = c_p \,e^{\la_p F}\,; \qquad \la_p = - (p-2)^{-1}\,, \quad
    c_p = [-\ga p \,(p-2)^2\,]^{\la_p} \,, \quad -\ga p >0 \,.
    \quad p \neq 0, 2 \,,
   \label{ad8}
  \ee
  that suggests the transformation $\psi = \psi_p \,z(\la_p \,F)$. Then we can replace Eq.(\ref{a.24}) for $\psi$ by the autonomous equation for $z$ or, equivalently, by the first-order system for $z$ and
  $y \equiv \pz \,$:
  \be
   \ppz + \pz + z - z^{p-1} = 0 \,,\qquad
   (z^{p-1} - z - 2\,y) \,dy = y \,dz \,.
   \label{ad9}
  \ee
   Although in general these equations cannot be solved analytically we may regard them almost integrable, at least for rational $p$. Indeed, the behavior of their solutions was analyzed in great detail, including exact asymptotic behavior. More recently, a few significant results were obtained on classical integrability of Eq.(\ref{ad9}), see \cite{Ber}, \cite{EFint}.

    Before turning to a more general approach for searching additional
  integrals, we mention an interesting integrable potential
  $U = g\xi + \bg \psi^2$. Then $F$ satisfies Eqs.(\ref{ad3}) and thus
  the potential $e^F$ is given by Eq.(\ref{ad5}); the linear  equation for  $\psi$ (see (\ref{ad6})) is related to the Legendre equation and can be explicitly solved. Moreover, for certain discrete values of the parameters  the potential in the equations for $\psi$ and $\xi$, which  is proportional to $\cosh^{-2} (c\tau)$, becomes `transparent'. Then the solution can be expressed in terms of elementary functions \cite{FI}.

  \section{On a more systematic approach to search of integrals}
   In our general approach to the DG dynamical systems we try to take care  for nice features both of the Hamiltonian and LC formulations. In  \cite{ATF1} and \cite{FM} we proposed a first-order dynamical system which is related to the Hamilton dynamics but does not coincide with it. This allowed us to find rather nontrivial integrals, to study some global properties of the solutions, and to construct convergent analytic expansions near horizons and singularities. Our method can be applied to many scalars but here we mostly consider the models with one scalar having arbitrary coupling $Z(\f)$ and arbitrary potential $V(\f, \psi)$.

   Introducing the new momentum-like variables  $\chi, \eta, \rho$,\footnote{
   It is easy to check $\chi = -p_F$, $\eta = -p_{\psi}/2$,
   $\rho = -Z p_{\f}$, where the momenta are derived for the Lagrangian (\ref{a.7a}) with LC gauge $l_{\ep} \,= 1$.   }
   \be
  \qquad    \quad \dot{\varphi} = \chi \,, \quad
 Z(\f)\, \dpsi = \eta \,, \quad Z(\f) \,\dF = \, \rho \,,
 \label{a.27}
 \ee
 we rewrite the main dynamical equations  and the constraint (\ref{a.7b}) in the form
 \be
   \dchi +  hV  = 0\,, \quad
  2 \deta + hV_{\psi} = 0 \,, \quad
  \dot{\rho} +h (ZV)_{\f} = 0\,; \qquad
  \chi \rho + h ZV + \eta^2 = 0\,.
  \label{a.28}
  \ee
    The equations (\ref{a.27}) and the first three equations of (\ref{a.28}) are equivalent to the canonical system with the Hamiltonian which is equal to the constraint divided by $Z$.
   To completely solve this system we must search for two additional constraints canonically commuting with the Hamiltonian and with each other.  As far as we are interested in the classical theory we usually will look for integrability in the Liouville sense. Moreover, we are mostly interested in explicit analytic expressions for the solutions or, at least, in  exact analytic relations between the `physical' variables
   $h, \f, \psi$.

   Looking at the system (\ref{a.28}) we immediately see that there exist integrals $\eta = \eta_0$ if $V_{\psi} = 0$ and $\rho = \rho_0$ if $(ZV)_{\f} = 0$. The best studied is the first case of the $\psi$-independent potential. When  $\eta_0 = 0$ the dilaton gravity can be explicitly solved with arbitrary potentials;
   the simplest case
   $(ZV)_{\f} = (ZV)_{\psi} = 0$ was solved above. In \cite{ATF1}, we derived three dilaton gravity models with $\eta = \eta_0\,$, for which there exists one more integral, and demonstrated how they can be explicitly solved. Two of them are closely related to the Liouville theory but the third one requires a generalization of the Liouville integral that will be discussed in a moment.

  Recalling the previous section, we change the variable $\tau$ to
  $\xi$, what is defined by the relations $\chi d\tau = d\f \equiv Z d\xi$. Then we rewrite equations
  (\ref{a.27})-(\ref{a.28}) denoting the derivative $d/d\xi$ by the prime  and introducing useful notation:
     $H(\xi) \equiv \, h/\chi\,$, $G(\xi) \equiv \,\eta/\chi \,$,
    and  $U \equiv Z V\,$ as above.
   The main independent equations for $\chi$, $\eta$, $\psi$, $H$ now have a very compact form:
  \be
  \prpsi = G \,, \quad \prH = -G^2 H\,, \qquad
  \prchi +\, U H = 0\,, \quad  2\preta +\, U_{\psi} H = 0\,,
     \label{a.29}
  \ee
   where the second equation is in fact equivalent to the constraint.
  The \emph{extended system} contains two equations with $\rho$
  (see (\ref{a.27})-(\ref{a.28})):
    \be
  \chi \prF - \rho = 0\,, \quad
  \prho + U_{\xi} H = 0 \,; \qquad
  \chi \prG = UH(G - U_{\psi} /2U)\,,
  \label{a.30}
  \ee
  where we add the explicit equation for $G$,
 which can replace the last equation in (\ref{a.29}).

    The system (\ref{a.29}) is most convenient for deriving the \emph{solutions near horizons} and in asymptotic regions as well as for studying their general properties.
    For example, a very important property  of its solutions is that
   $(\ln H)^{\pr} = - G^2 <0$.  This property does not depend on the potential and is true for any number of scalar fields if their  $Z$-functions are negative,  as was first shown in \cite{ATFp}.
   Indeed, in this case the constraint equation can be written as:
   \be
   \prPhi \equiv \,(\ln H)^{\pr} =
   - Z_0 \sum_{n=0}^N Z^{-1}_n (\xi)\, (\eta_n /\chi)^2   \,, \qquad
   \eta_n \equiv Z_n \, \dpsi_n \equiv \chi Z_n Z_0^{-1}\, \prpsi_n \,.
  \label{a.30a}
  \ee
   For normal fields $Z_n < 0$ for all $n$, and thus $\prPhi <0$.
   For anomalous fields,
   like the scalaron corresponding to the tachyonic vecton,
    $Z$ may be positive and then the sign of $\prPhi$  may be negative or positive, depending on  concrete solutions. In case of the same signs, Eq.(\ref{a.30a}) resembles the second law of thermodynamics and defines an \emph{`arrow of time'} for our dynamical system. If this is not true, the theorem is violated in a very specific way. It may be an interesting point for  cosmological modeling.

   Our system of equations (\ref{a.29}) has other interesting global properties. The general solution of the first two equations can be written in terms of integrals of $G(\xi)$ and $G^2(\xi)$:
     \be
  \psi (\xi) \,=\,  \psi_0 \,+\, \int_{\xi_0}^\xi G \,,   \qquad
  H(\xi) \,=\, H_0 \exp \biggl(-\int_{\xi_0}^\xi G^{\,2} \biggr) \,.
    \label{a.31}
  \ee
   Then, inserting these `solutions' into the third and the forth equations and integrating them we can write one integral equation for  $G(\xi)$ instead of  system (\ref{a.29}):
      \be
  G (\xi) \equiv {\eta \over \chi} \,=\,  \biggl(\,\eta_0 \,-\,
  \half \, \int_{\xi_0}^\xi U_{\psi} H \biggr) \,
   \biggl( \,\chi_0  \,-\,   \int_{\xi_0}^\xi U H \biggr)^{-1} \,,
    \label{a.32}
  \ee
   where $\psi(\xi)$ and $H(\xi)$ are given by Eq.(\ref{a.31}). As was discussed in \cite{ATF1} and \cite{FM}, the standard
   (regular and non-degenerate) horizon appears when
   $\chi_0 = \,\eta_0 = 0$. Then $h(\xi_0) = 0$ while
   $G(\xi_0) = U_{\psi} (\xi_0)/\,2U (\xi_0)$ is finite if
   $U_{\psi} \neq 0$. It follows that $G$, $\psi$, $H$ are finite and can be expanded in convergent series around $\xi_0$ if the potential
   $U(\xi, \psi)$ is analytic in a neighborhood of $(\xi_0, \psi_0)$.\footnote{
   This is shown in \cite{FM}.  One can find a detailed discussion of regular solution with horizons, including a generalization of the Szekeres - Kruskal coordinates and examples of singular horizons, in \cite{ATFp}.}
   When $U_{\psi} \equiv 0$ there is the obvious integral of motion
   $\eta = \eta_0$. As can be seen from the above equations and was proved in  \cite{ATF1}, there is no horizon if $\eta_0 \neq0$; when $\eta_0 = 0$, we have $G \equiv 0$ and return to
   pure dilaton gravity with horizons.

   In simple cases, the integral equation can easily be reduced to a differential one. For example, if $U = u(\xi)\, v(\psi)$ and $U_{\psi} = 2g \,U$, the integral equation can be reduced to the second-order
   differential equation, which is not integrable for arbitrary $u(\xi)$
   but is explicitly integrable if $U_\xi = g_1 U$. This result is quite natural as in this case there exist two additional integrals, 
   $\eta = g \chi + \eta_0\,$, $\rho = g_1 \chi + \rho_0\,$, and therefore
   the most direct approach is to use the extended differential system.
   In non-integrable cases or when there is only one additional integral, this \emph{`master' integral equation} still can be a quite useful analytical tool.

   Above we mostly supposed that the potential $U$ is known and tried to find integrals or directly integrate some equations. Now we consider a different approach supposing that we do not fix the potential and try to find equations for the potentials allowing some integrals. To get a feeling of the approach look at Eq.(\ref{ad2}). If we take potentials, for which the expression in brackets vanishes, we immediately find that the solution of the homogeneous equation for $\ln U$ depends on an arbitrary function of one variable,  $f(c_2 \xi - 2 c_3 \psi)$, and on arbitrary parameter defining solutions of inhomogeneous equation. In general, we thus obtain one integral. However, for linear $f$ we can derive two independent integrals. This approach becomes much more powerful if we use  extended system (\ref{a.29}), (\ref{a.30}).
      Here, we only briefly outline a generalization of the approach of Ref.\cite{ATF1} to finding potentials $U$ for which the extended differential system has additional integrals (see also \cite{DF}).

  Generalizing the approach of \cite{ATF1} and the above remarks
  about possible integrals of motion let us collect those equations which can generate such integrals:
   \be
  0 \,=\, \rho \,+\, U H \,+\, \eta^2 /\chi \,= \,
   \prchi +\, U H \,=\,  \preta +\, U_{\psi} \,H/2 \,=\,
    \prho + U_{\xi} H \,= \,0\,,
  \label{a.33}
  \ee
  \be
  0 \,=\, \psi \,\preta +\, \psi \,U_{\psi}\, H/2 \,=\,
  \eta \,\prpsi - \eta^2 /\chi \,= \,
   \xi \,\prho + \xi\, U_{\xi}\, H \,= \,
   \xi^{\pr} \rho \,-\, \rho \,= \,0 \,.
    \label{a.33a}
  \ee
  The first equation in (\ref{a.33}) is the energy constraint, which we denote  $E_0$ and the next items in this chain of equations
  are denoted by $E_i\,$, $i = 1,...,7$. Now, taking the sum
  $\sum_0^7 c_i E_i$ with $c_4 = c_5 =c_6 =c_7 = c_0$ we find that
  the solutions of equations (\ref{a.33}) satisfy the identity
   \be
   [c_1 \chi +\, c_2 \,\eta +\, c_3 \,\rho +\, c_4 (\psi \,\eta +\, \xi \,\rho)]^{\pr} =
   -H [(c_1 +\, c_4)\,U +\, c_2 \,U_{\psi}/2 +\, c_3 \,U_{\xi} +\,
   c_4 (\psi \,U_{\psi}/2 + \xi \,U_{\xi})]\,.
  \label{a.34}
  \ee
    Therefore, if the r.h.s. identically vanishes, the l.h.s generates the integral of motion,
   \be
   c_1 \chi +\, c_2 \,\eta +\, c_3 \,\rho +\, c_4 (\psi \,\eta +\, \xi \,\rho) = I \,.
   \label{a.35}
   \ee
    This means that for the potentials $U(\xi, \psi)$, satisfying the partial differential equation
   \be
   (c_1 +\, c_4)\,U +\, c_2 \,U_{\psi}/2 +\, c_3 \,U_{\xi} +\,
   c_4 (\psi \,U_{\psi}/2 + \xi \,U_{\xi})\,=\,0 \,,
  \label{a.36}
  \ee
  there exist the corresponding integral of equations (\ref{a.33}). Many of  the  above   integrals can be obtained by applying this theorem.
    The solution of Eq.(\ref{a.36}) depends on an arbitrary function of one variable. Using this fact  it is possible, in some simple cases, to derive one more integral. This is true, for example if the potential is exponential.

   In the above example of the potential $U = U_0 \exp(2g\psi + g_1\xi)$
   with two additional integrals we immediately find the equation for $\chi$,
   \be
   \prchi = (g^2 + \,g_1)\, \chi \,+\, (\rho_0 + 2\, \eta_0 \,g)\, +\,
   \eta_0^2/\chi \,,
   \label{a.37}
   \ee
   by solving of which we explicitly express $\xi$, $\eta$, $\rho$, $h$ and $\psi$ as functions of $\chi$. This is sufficient for finding the portrait of this physically interesting system.

  To demonstrate the problems, which remain even for apparently simpler systems, we consider the potential $U(\psi)$ also having two additional integrals. The obvious linear integral is $\rho = \rho_0$. To obtain one more integral suppose that $\psi U_{\psi} = 2 g U$. Then we have the additional integral of the three differential equations:
    \be
    \psi \eta -(g +1)\,\chi + \rho_0 \,\xi = I\,; \qquad
    \prchi = \rho_0 + \eta^2/\chi\,, \quad
    \eta \,\prpsi = \eta^2 /\chi  \,, \quad \psi \,\preta = g \prchi \,.
   \label{a.38}
   \ee
   We can exclude $\psi$ (or, $\eta$) and thus get two equation for $\chi$ and $\eta$  (or $\psi$).  But this is not an integrable dynamical system because of its explicit dependence of $I$ on $\xi$. If we take $\rho_0 = 0$ in the expression for the above integral, the system can be explicitly integrated but this is only a \emph{`partial' solution}.\footnote{This is a typical problem -- integrals often have $\xi$-depending terms that describe a sort of a `back-reaction' of gravity on matter.} Unfortunately, here we do not know the transformation to an autonomous system, which helped us in the Emden-Fowler case, see (\ref{ad8}), (\ref{ad9}).

   It should be emphasized that equations (\ref{a.34})-(\ref{a.36}) only generate  the integrals that are linear in momenta (like the variables
   $\chi, \eta, \rho$), but do not allow derivation of possible bilinear integrals.
     The simplest example of an additional \emph{bilinear integral}  can be obtained if we multiply the last equation in (\ref{a.33}) by $2\rho$,
   \be
     0 = 2 \rho \,(\rho^{\pr} + \,U_{\xi} H ) \equiv
       \,(\rho^2)^{\pr}  + 2 h^{\pr} U_{\xi}   \,,
       \label{a.39}
   \ee
   and then suppose that $U_{\xi}$ is a constant, $U_{\xi} = g$.
    This gives a new integral, $\rho^2 + 2 g h  = I$, generalizing the Liouville one. This integral exists if $U = g \,\xi + v(\psi)$ and we can find $v(\psi)$, for which there exists one more integral, with the aid of equation (\ref{a.36}). Inserting in it the expression for $U$ we find that $c_1 +2 c_4 = 0$ and thus find the equation for $v$:
    \be
    (c_2 + c_4  \,\psi)\,v^{\pr}(\psi) - 2\,c_4 v(\psi) + 2\,g \,c_3
    \,=\,0 \,.
    \ee
     If $c_4 =0$ the solutions is linear in $\psi$ and we have the additional integral $c_2 \,\eta + c_3 \,\rho = I$. If $c_4 \neq 0$, we find that $v = g\,c_3/c_4 +c_0 (\psi + c_2/c_4)^2$ and the integral is
     \be
     -2 c_4 \chi +c_2 \eta + + c_3 \rho + c_4 (\psi \eta + \rho \xi)
     = I \,.
     \ee

    Some integrals of this sort were first discovered in \cite{ATF1}. To derive them using the present approach we somewhat generalize this process and find the following nontrivial integral:
    \be
    (a + b\,\xi)\,\rho^2 +c h + b \,\eta_0^2 \,\ln |h| = I \,,
    \label{a.39a}
    \ee
    where we used the constraint multiplied by $\rho$, the identity
    $\rho H = \prh U$, and supposed that $U_{\psi} = 0$, which gives the integral  $\eta = \eta_0\,$.  The solution of the equation for the potential is
    \bdm
    U = {c \over b} \,+\, {c_0 \over \sqrt{a + b \,\xi}} \,,
    \edm
    where $c_0$ is arbitrary. The integral and potential do not coincide with those of Ref.\cite{ATF1} and look rather exotic. To reproduce the most interesting integral of \cite{ATF1} we note that here we work exclusively in Weyl's frame $W = 0$ (but omitting hats) while there was used the original frame $W = (1-\nu)/ \f$, see equations (\ref{2.2}), (\ref{2.4}), (\ref{2.5}) above.

        To return to the general W-frame with the arbitrary dilaton kinetic potential $W(\f) \equiv W(\xi)$, we thus apply the Weyl transformation $h \mapsto hw$, $V \mapsto V/w$ and, correspondingly,
    \be
    H \mapsto Hw, \quad U \mapsto U / w, \quad F \mapsto F + \ln w,  \quad \rho \mapsto \rho + \chi (\ln w)^\pr.
    \label{a.40}
    \ee
    This gives us more freedom in search for integrals. For example,
    if we apply (\ref{a.40}) to (\ref{a.39}), replace $h$, $U$, $\rho$ by $\hh \equiv h w$, $\hU \equiv U/w$, $\hrho$, transform $\hh^{\pr}\, \hU_{\xi}$ into
    \bdm
      h^{\pr}\, [w\, \hU_{\xi}] \,+\, h U\, [U^{-1} w^{\pr}\, \hU_{\xi}] \,=\,  h^{\pr}\, [w\, \hU_{\xi}] \,-\, \chi \prchi \,[U^{-1} w^{\pr}\, \hU_{\xi}] \,,
    \edm
    and make the expressions in the square brackets constant, we will find a new bilinear integral. With this approach, we can recover the complex bilinear integral of \cite{ATF1} and find new ones. This subject requires more careful investigations and will be discussed elsewhere.

  Finally, let us return to our main goal formulated in first and second Sections -- to integrate the equations or, at least, to find a global portrait of the simplest or approximate scalaron cosmologies. Here we present a realistic example of one additional integral. Consider the integral and the corresponding equation for the potential
  \be
  c_1 \chi + (\psi \eta + \rho \xi) = I \,, \quad
  2 \beta \,U + 2 \,\xi \,U_{\xi} + \psi \,U_\psi = 0 \,; \qquad
  \beta \equiv c_1 + 1 \,.
  \ee
  The general solution for this equation is
  \be
   U (\xi, \psi) = \psi^{2\al}\, \xi^{\beta - \al} \,
   \cF (1 + \psi^2 \,\xi^{-1}) \,,
  \ee
  where $\cF(x)$ is an arbitrary function of one variable, $\alpha$ --  arbitrary parameter. Taking
  \bdm
  \cF \equiv \cF_1 = (1 + \la_0^2  \,\psi^2 /\xi) \,, \qquad \al =0\,, \,\,\,\,
  \beta = 0 \,,
  \edm
  we get the effective potential for the scalaron model in $D = 3$, see (\ref{3.14}), (\ref{a.15}). For other dimensions, neglecting the the curvature term (i.e., $k_{\nu} = 0$), Eq.(\ref{a.15}) gives
  \bdm
  \cF \equiv \cF_{\nu} = v_{\nu} (\la_0 \,\psi /\f) \,; \qquad
  \al = 0 \,, \quad \beta = (\nu-1)/2 \,.
  \edm
  We still hope to find one more integral for $D = 3$. However, for other dimensions we must look for some approximations.

 \section{Summary and outlook}
    In conclusion we  summarize the main points of the report.
    Dilaton gravity with scalars is in general not integrable even with formally sufficient number of integrals of motion. The models with massless scalars qualitatively differ from the scalaron models (DSG) that inevitably include non-integrability. Fortunately, in some physically important cases the non-integrable systems are partially integrable and therefore can be effectively studied, at least qualitatively. The solutions near horizons and singularities can be
     derived analytically -- by using exact series expansions or, alternatively, by iterations of the master integral equation. On the other hand, our approach to constructing systems with additional integrals may help to find integrable  or partially integrable systems that are qualitatively close to the realistic ones.
     We demonstrated that there are two kinds of additional integrals: 1. linear in momenta and described by a sufficiently general approach; 2. quadratic in momenta, similar to the Liouville integrals. In addition, we argued that in the dilaton gravity the scalar field equations resemble the well known generalized Emden-Fowler equations. The rich technology developed for understanding the global structure and asymptotic properties of these equations may prove helpful in the context of scalaron models in cosmology.

     Understanding global properties of classical solutions is also desirable for their quantization. The simplest approach was attempted some time ago for classically integrable gravitational systems with minimal number of degrees of freedom (see, e.g.  \cite{CAF}).

     We hope that the above panoramic presentation of several new ideas on   finding integrals of nonlinear equations of modern cosmological models, which are met in various generalized theories of gravity,
      may be of interest in studies of their global properties.
    
  \bigskip
   {\bf Acknowledgment}
  Useful remarks of E.A.~Davydov are kindly acknowledged.

 %\newpage

 %\bigskip
  
 \end{document}